\documentclass[twocolumn,showpacs,preprintnumbers,amsmath,amssymb]{revtex4}
\pdfoutput=1

\usepackage{graphicx}
\usepackage{dcolumn}
\usepackage{bm}

\begin{document}

\title{Spin-induced anomalous magnetoresistance at the (100) surface of hydrogen-terminated diamond}
\author{Yamaguchi Takahide$^{1,2}$, Yosuke Sasama$^{1,2}$, Masashi Tanaka$^{1}$\\ Hiroyuki Takeya$^{1}$, Yoshihiko Takano$^{1,2}$, Taisuke Kageura$^{3}$, Hiroshi Kawarada$^{3}$}
\affiliation{$^{1}$National Institute for Materials Science, Sengen, Tsukuba 305-0047, Japan\\
$^{2}$University of Tsukuba, Tennodai, Tsukuba, 305-8571, Japan\\
$^{3}$Waseda University, Okubo, Shinjuku, Tokyo 169-8555, Japan\\}
\date{\today}

\begin{abstract}

We report magnetoresistance measurements of hydrogen-terminated (100)-oriented diamond surfaces where hole carriers are accumulated using an ionic-liquid-gated field-effect-transistor technique. Unexpectedly, the observed magnetoresistance is positive within the range of $2{\textless}T{\textless}10$ K and $-7{\textless}B{\textless}7$ T, in striking contrast to the negative magnetoresistance previously detected for similar devices with (111)-oriented diamond surfaces. Furthermore we find: 1) magnetoresistance is orders of magnitude larger than that of the classical orbital magnetoresistance; 2) magnetoresistance is nearly independent of the direction of the applied magnetic field; 3) for the in-plane field, the magnetoresistance ratio defined as $[\rho(B)-\rho(0)]/\rho(0)$ follows a universal function of $B/T$. These results indicate that the spin degree of freedom of hole carriers plays an important role in the surface conductivity of hydrogen-terminated (100) diamond.
\end{abstract}



\maketitle


Electronic spins at nitrogen-vacancy centers and other defects in diamond have attracted intense interest mainly because of their long coherence times, appropriate for applications in quantum information processing and sensitive magnetometry\cite{Doh13,Ron14}. In contrast to such localized spins, studies on the spins of conductive electrons and holes in diamond have been limited, although they are also expected to have long coherence times due to the weak spin-orbit coupling\cite{Res12}, which promises spintronics applications. This is partly due to the difficulty in inducing a high conductivity in diamond: the n-type conductivity, induced by phosphorous doping, is very low\cite{Kat04,Kat09}, and the p-type conductivity also requires a much heavier doping of boron than in the case of silicon\cite{Bor96}. Obtaining a moderate conductivity using the field effect doping also requires a large sheet carrier density above $10^{13}$ cm$^{-2}$\cite{Yam13}. 

Recently, the control of such a large carrier density has become possible\cite{Yam13,Yam14} with the use of an ionic-liquid-gated field-effect-transistor technique\cite{Fuj13,Uen14}, which employs a large capacitance of the electric double layer formed on the channel surface. Naturally, the condition of the channel surface is crucial in this doping technique. As diamond has a three-dimensional covalent crystal structure, we have unique options regarding the surface preparation; that is, we can choose the surface crystal orientation and the kind of atoms that terminate the surface dangling bonds. These options, which are not expected in layered compounds like transition metal dichalcogenides\cite{Wan12}, may provide additional novel properties and functionalities for the transport of the accumulated carriers.

In this paper, we report an anomalous positive magnetoresistance effect induced by the spins of conductive holes at the surface of hydrogen-terminated (100)-oriented diamond. Hole carriers with the density above $10^{13}$ cm$^{-2}$ are accumulated at the diamond surface using an ionic liquid gating, which enables systematic transport measurements at low temperatures. Interestingly, the positive magnetoresistance for the (100) surface contrasts with the negative magnetoresistance for the (111) surface\cite{Yam14}. The magnetoresistance is orders of magnitude larger than that of the classical orbital magnetoresistance and is nearly independent of the magnetic field orientation. We also find that the in-plane magnetoresistance curves at different temperatures collapse onto a single curve when they are plotted as a function of $B/T$. These results indicate that the spins of the conductive holes play an essential role at the (100) diamond surface, which may have implications for the development of diamond-based spintronics.

We fabricated ionic-liquid-gated field effect transistors on hydrogen-terminated (100)-oriented IIa-type single crystal diamonds (Fig. 1(a)). The hydrogen termination raises the energy bands of diamond relative to the vacuum level, thus favoring the introduction of hole carriers.\cite{Squ06} Even the exposure of the hydrogen-terminated surface to the air introduces hole carriers, which will probably be due to the electron transfer from the top of the valence band to the redox level of H$_3$O$^{+}$/H$_2$ in an adsorbed water layer.\cite{Squ06,Lan89,Kaw96,Mai00,Neb04} A Hall bar, used as the channel of the transistor, was produced using photolithography and a UV ozone treatment. After heating the sample in Ar atmosphere to reduce the density of adsorbates on the channel surface, a small amount of ionic liquid, N,N-diethyl-N-methyl-N-(2-methoxyethl)ammonium bis(trifluoromethylsulfonyl)imide (DEME-TFSI; Kanto Chemical), was applied between the channel and a gate electrode. Details of the device fabrication are described in Supplementary Material (SM). The transport measurements were performed in a custom-built cryostat probe inserted in a physical property measurement system (Quantum Design). The resistance and Hall voltage were measured using voltage and current preamplifiers in an ohmic region with a current less than 10 nA.

\begin{figure}
\includegraphics[width=8.5truecm]{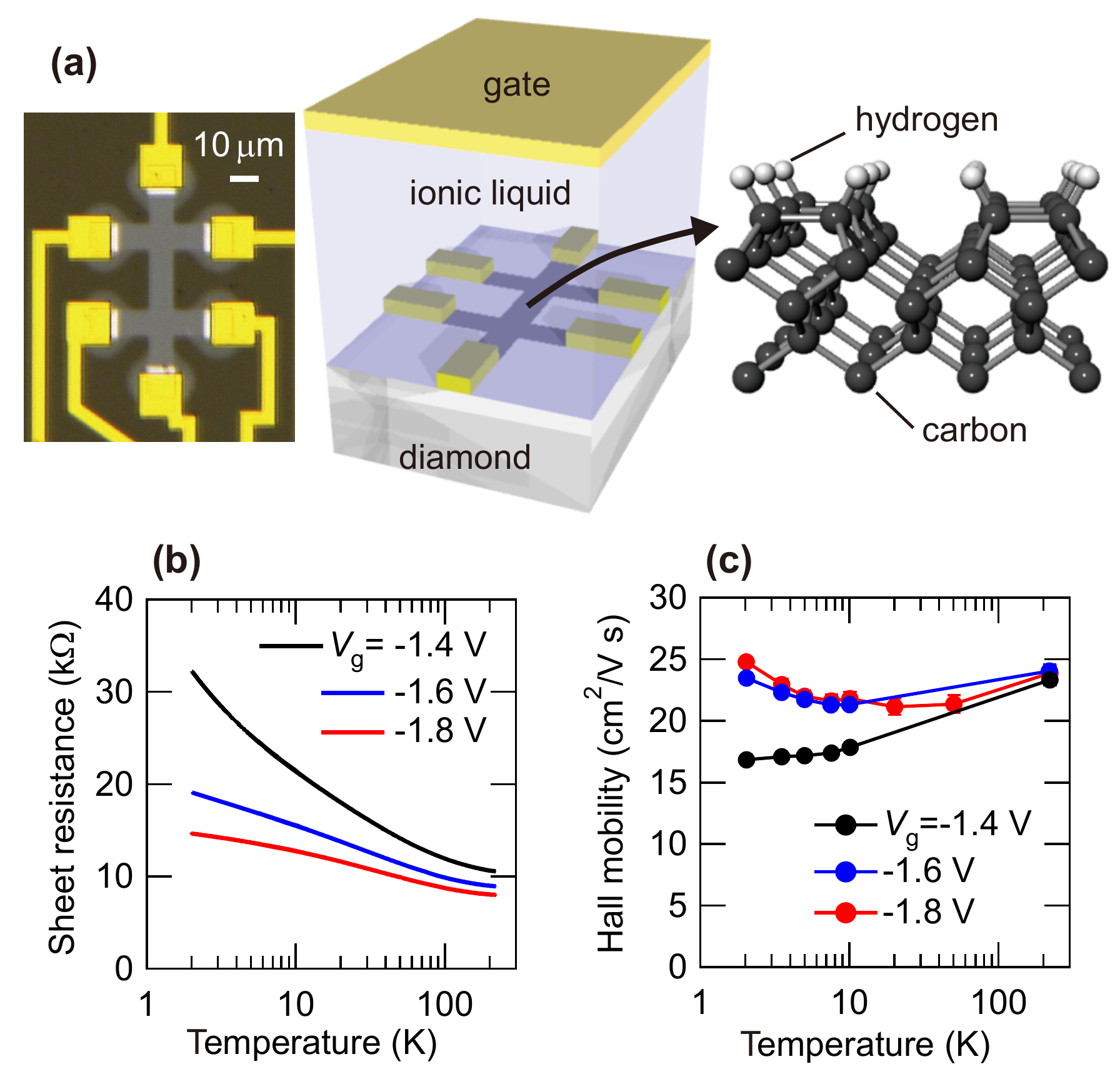}
\caption{(a) An optical micrograph of the Hall bar and a schematic diagram of an ionic-liquid-gated field effect transistor on the hydrogen-terminated (100) diamond surface. (b, c) Temperature dependences of the sheet resistance $\rho$ (b) and Hall mobility (c) for the gate voltages $V_g=-1.4$, -1.6 and -1.8 V.}
\end{figure}

To accumulate hole carriers at the diamond surface, we applied negative voltage to the gate at 220 K, slightly higher than the glass transition temperature of the ionic liquid. The gate voltage dependence of the channel resistance was nearly reversible when the gate voltage was less than 1.8 V at 220 K, indicating an electrostatic accumulation of carriers. The temperature dependences of the sheet resistance and Hall mobility at three different gate voltages are shown in Figs. 1(b) and (c). Here we focus on the gate voltage regime for which the temperature dependence of resistance is close to log($T$) at low temperature. At a lower gate voltage, the resistance shows an activated temperature dependence.\cite{Yam13} The Hall carrier density at 2 K was $1.15{\times}10^{13}$, $1.39{\times}10^{13}$, and $1.72\times10^{13}$ cm$^{-2}$ for $V_g=-1.4$, $-1.6$, and $-1.8$ V, respectively. (See Fig. S1 in SM for the temperature dependences of the Hall coefficient $\rho_{xy}/B$ and Hall carrier density)

\begin{figure}
\includegraphics[width=5.3truecm]{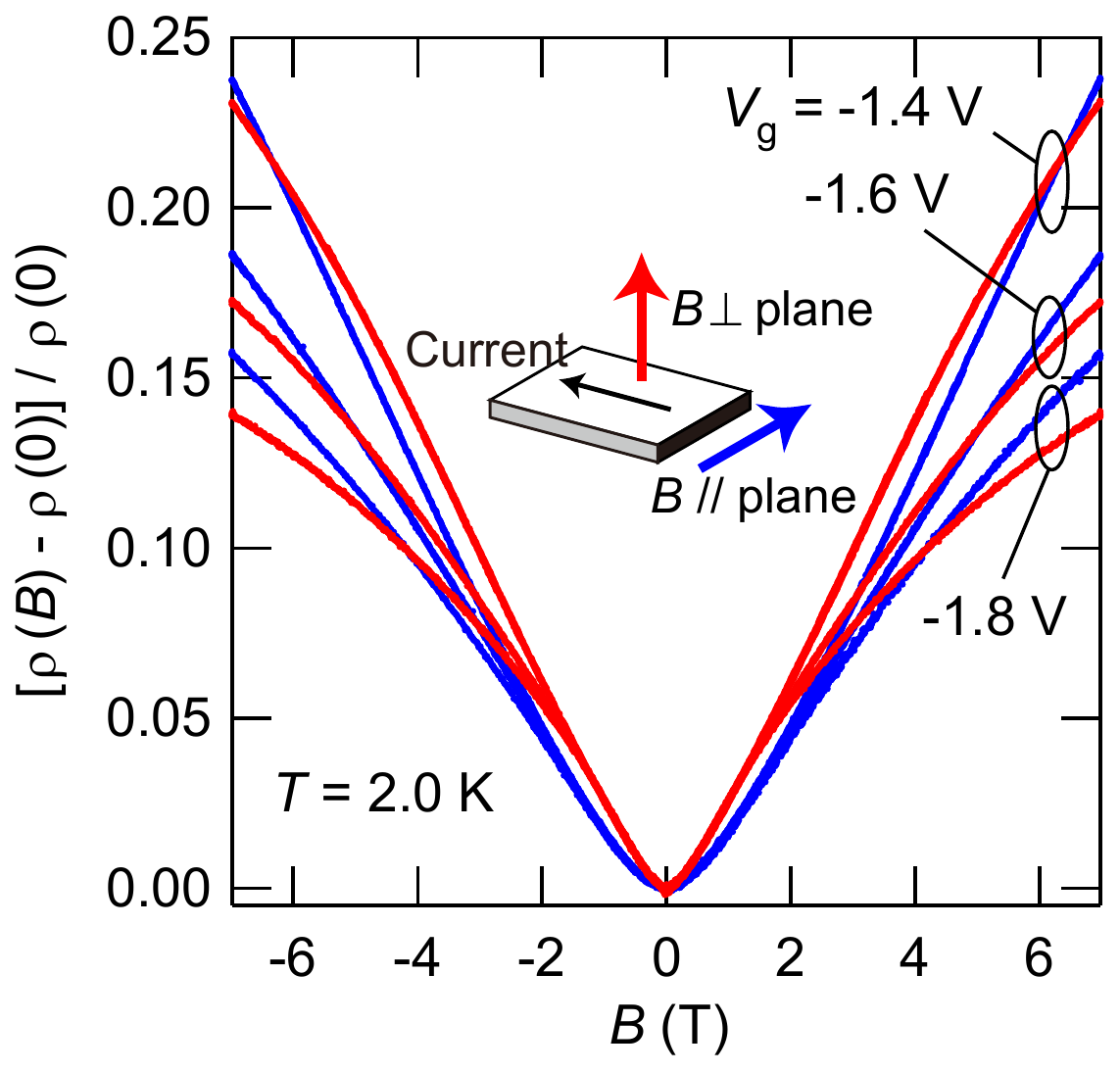}
\caption{Magnetic field ($B$) dependence of the magnetoresistance ratio $[\rho(B)-\rho(0)]/\rho(0)$ measured for $V_g=-1.4$, -1.6 and -1.8 V at $T= 2.0$ K by applying a magnetic field parallel and perpendicular to the diamond surface.}
\end{figure}

We measured the magnetoresistance at 2 K for each gate voltage. Figure 2 shows the magnetoresistance ratio $[\rho(B)-\rho(0)]/\rho(0)$ as a function of the magnetic field $B$. Here, the magnetic field is applied parallel and perpendicular to the diamond surface. This figure shows some distinctive features of the magnetoresistance of (100) diamond surface. First, the magnetoresistance is positive; namely, the resistance increases with increasing magnetic field. This is in striking contrast to the negative magnetoresistance observed for similar samples with (111) diamond surfaces, the latter being attributed to two-dimensional weak localization\cite{Yam14} (See also Sec. 5 of SM). Furthermore, the magnetoresistance is orders of magnitude larger than the positive, classical orbital magnetoresistance if one considers the Hall mobility of 17-25 cm$^2$/Vs (Fig. 1(c)). We note that the surface of the (100) diamond used here has a larger roughness (root-mean-square value $0.2-0.5$ nm for a 1 ${\mu}$m$^2$ area) than the (111) diamond used in Ref.\cite{Yam14}. However, this is not the origin of the difference in the magnetoresistance effect; the negative magnetoresistance has also been observed in (111) samples with the surface roughness comparable to that of the (100) samples in this study.

Another important feature for the (100) surface is that a large positive magnetoresistance is seen even in the case that the magnetic field is applied parallel to the surface, and its magnitude is comparable to that for the perpendicular field. Here, the magnetic field is parallel to the surface and perpendicular to the current. Approximately the same magnetoresistance curves are also obtained in the case that the field is parallel to the current. (Figs. S2 and S3 in SM) The in-plane field does not affect the orbital motion of the carriers. Therefore, the large in-plane magnetoresistance indicates that the spin degree of freedom of the carriers is important in the magnetotransport.

\begin{figure}
\includegraphics[width=8.5truecm]{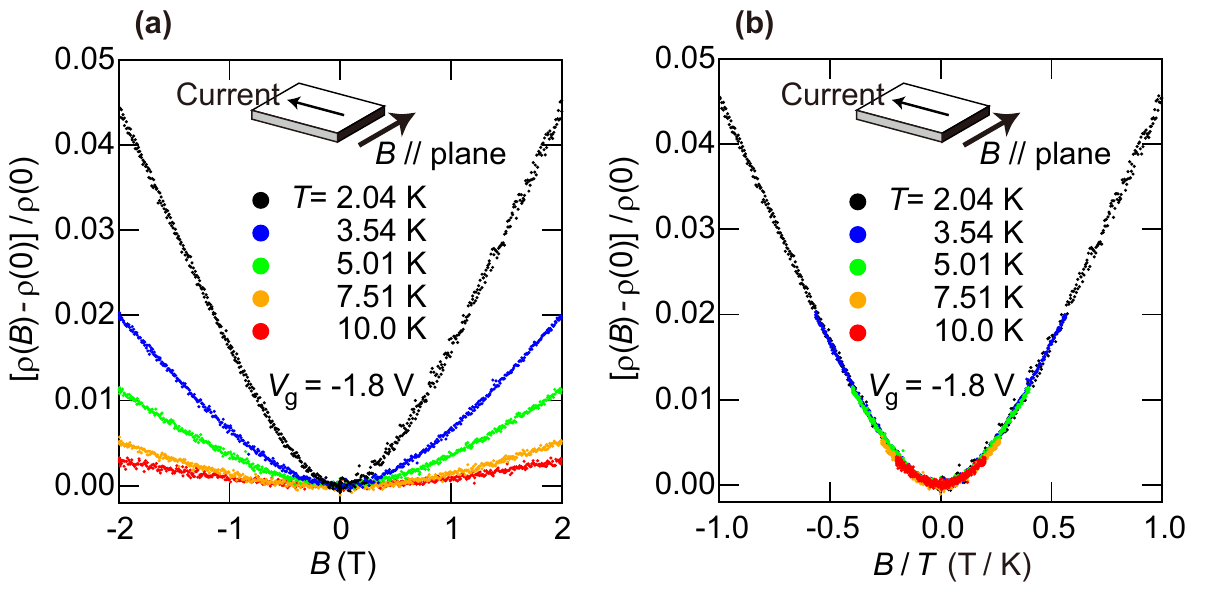}
\caption{(a) Magnetic field ($B$) dependence of the magnetoresistance ratio $[\rho(B)-\rho(0)]/\rho(0)$ measured for $V_g=-1.8$ V at different temperatures by applying a magnetic field parallel to the diamond surface. (b) Plots of $[\rho(B)-\rho(0)]/\rho(0)$ as a function of $B/T$.}
\end{figure}

To further investigate the magnetoresistance effect we measured its temperature dependence. Figure 3(a) shows the in-plane magnetoresistance ratio $[\rho(B)-\rho(0)]/\rho(0)$ for $V_g=-1.8$ V at different temperatures. The magnetoresistance strongly depends on temperature at $T{\le}10$ K, at which the Hall mobility does not depend on temperature very much (Fig. 1(c)). This also excludes the possibility of the classical orbital magnetoresistance. A notable feature, shown in Fig. 3(b), is that these magnetoresistance data collapse onto a single curve when they are plotted as a function of $B/T$. At small $B/T{\le}0.3$ (T/K), this curve is parabolic with respect to $B/T$: $[\sigma(B)-\sigma(0)]/\sigma(0){\approx}-0.083(B/T)^2$. Contrastingly, the out-of-plane magnetoresistance data (Fig. 4 (a)) do not scale with $B/T$ as shown in Fig. 4 (b). We find instead that they scale with $B/T^{1.32}$ (Fig. 4 (c)). The exponent 1.32 was determined to minimize the mean square difference between the data at different temperatures. The exponent is 1.28 and 1.26 for $V_g=-1.4$ and -1.6 V, respectively, and 1.28 for $V_g=-1.0$ V for another sample. These results may suggest that the out-of-plane magnetoresistance scales with $B/T^{4/3}$ or $B/T^{5/4}$.

\begin{figure}
\includegraphics[width=8.5truecm]{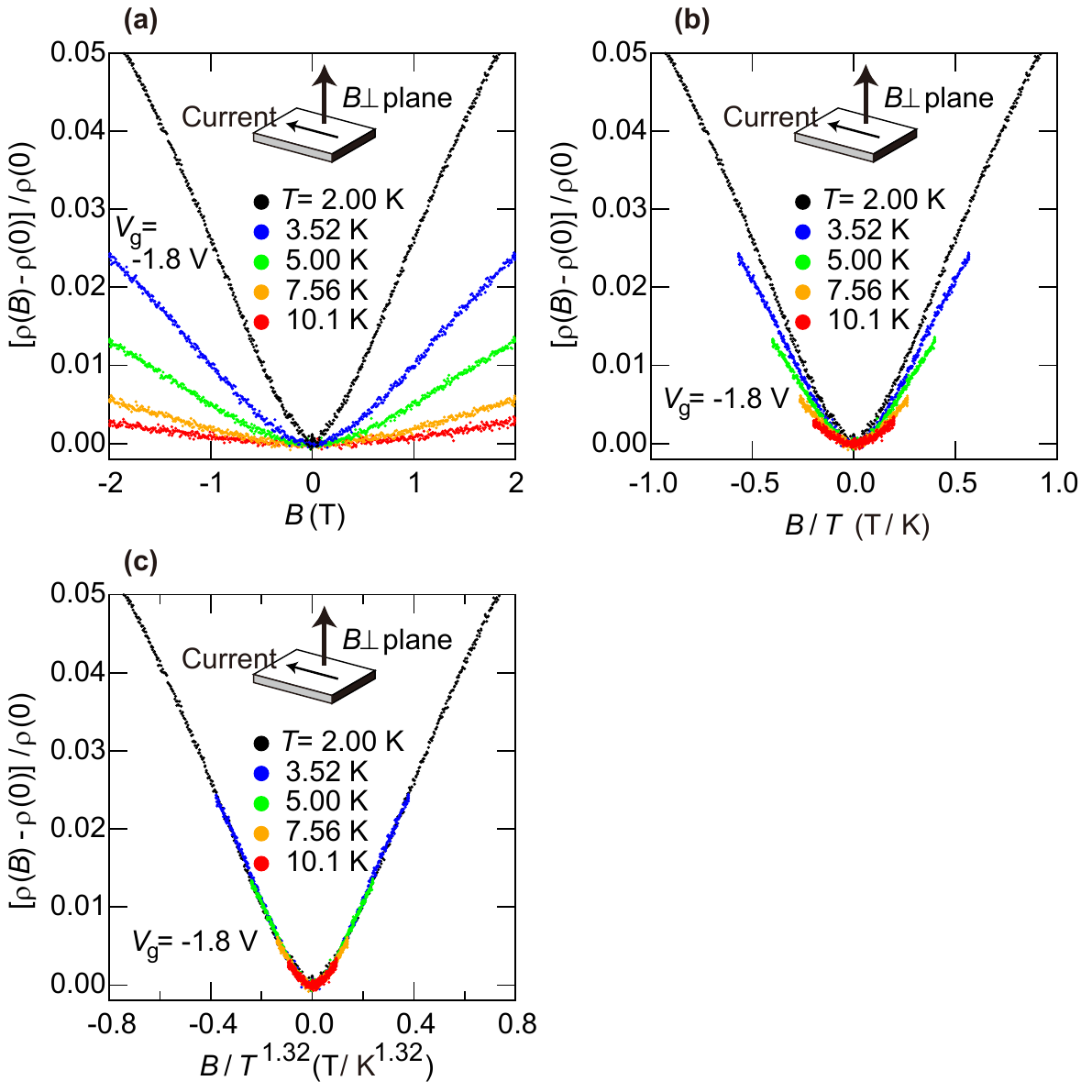}
\caption{(a) Magnetic field ($B$) dependence of the magnetoresistance ratio $[\rho(B)-\rho(0)]/\rho(0)$ measured for $V_g=-1.8$ V at different temperatures by applying a magnetic field perpendicular to the diamond surface. (b) Plots of $[\rho(B)-\rho(0)]/\rho(0)$ as a function of $B/T$. (c) Plots of $[\rho(B)-\rho(0)]/\rho(0)$ as a function of $B/T^{1.32}$.}
\end{figure}

We examine the scaling of the in-plane magnetoresistance in more detail below. Here we use the conductivity $\sigma(B)$ instead of the resistivity $\rho(B)$ for the purpose of comparison with the literature. Note that as $\rho_{xy}$ is much smaller than $\rho_{xx}$, $\sigma(B)$ is approximately the inverse of resistivity $\rho(B)$. Figure 5(a) shows plots of the in-plane $[\sigma(B)-\sigma(0)]/\sigma(0)$ at $V_g=-1.4, -1.6$, and -1.8 V as a function $B/T$. Not only the data at different temperatures at a certain gate voltage, but also the data at different gate voltages all collapse onto a single curve. Moreover, the data for another sample also collapse onto the same curve. (Fig. S3 in SM) On the other hand, a $B/T$ scaling of $\sigma(B)-\sigma(0)$, instead of $[\sigma(B)-\sigma(0)]/\sigma(0)$, has been reported in Ge bicrystals\cite{Rem84} and Si MOSFETs\cite{Sim98,Sim98_2}. The data shown in Fig. 5(a) are plotted in the form of $\sigma(B)-\sigma(0)$ vs. $B/T$ in Fig. 5(b). Obviously, the data at different gate voltages do not collapse onto a single curve. This indicates that $[\sigma(B)-\sigma(0)]/\sigma(0)$ rather than $\sigma(B)-\sigma(0)$ follows a universal function of $B/T$. Note that the data at each gate voltage appear to scale with $B/T$ to some degree in Fig. 5(b). This is because the temperature dependence of $\sigma(0)$ is not large. However, the deviation between 2.04 and 3.54 K is evident in Fig. 5(b) for $V_g=-1.4$ V, for which the temperature dependence of $\sigma(0)$ is the largest. This also indicates that $\sigma(B)-\sigma(0)$ is not a good quantity for the $B/T$ scaling. We presume that an appropriate quantity may be $[\sigma(B)-\sigma(0)]/\sigma(0)$ in Ge bicrystals and Si MOSFETs, too, because the reported $B/T$ scaling of $\sigma(B)-\sigma(0)$ was obtained at a certain gate voltage or for a certain carrier density. In fact, another form of scaling has also been reported in Si MOSFETs: $[(\sigma(0)-\sigma(B)]/[\sigma(0)-\sigma(\infty)] = f(B/T)$\cite{Vit01,Tsu05}, where $\sigma(\infty)$ is a temperature dependent constant. If $\sigma(\infty)$ is regarded as a field-independent part of conductivity, this scaling means that the remaining part $\sigma_r(B)\equiv\sigma(B)-\sigma(\infty)$ follows the scaling $[\sigma_r(B)-\sigma_r(0)]/\sigma_r(0) = -f(B/T)$.

\begin{figure}
\includegraphics[width=8.5truecm]{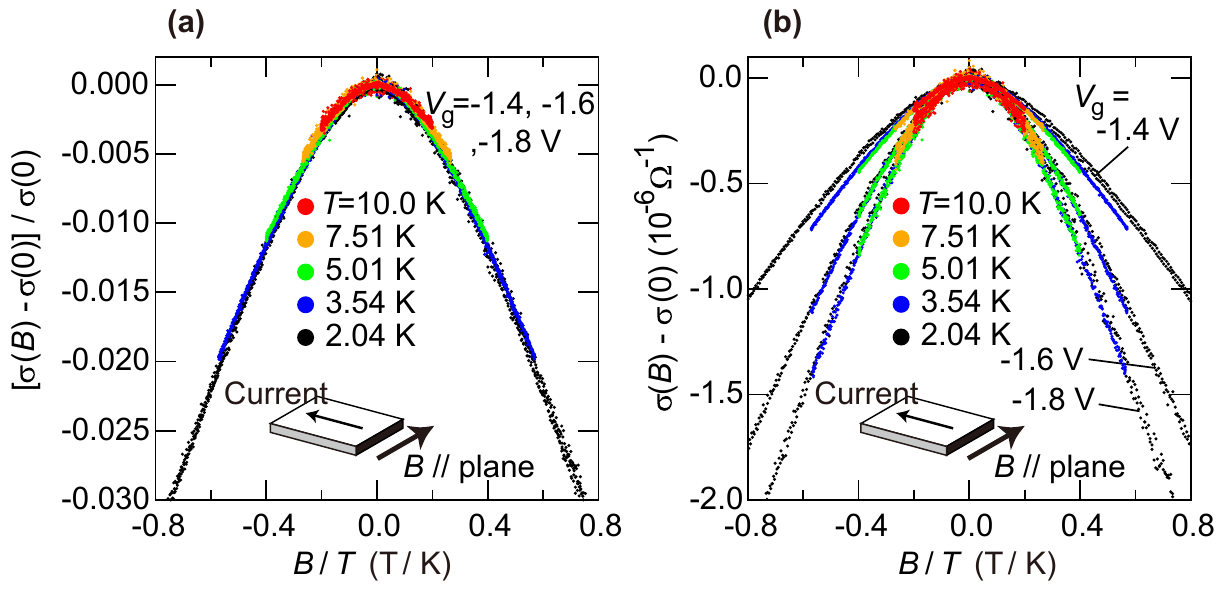}
\caption{ (a) Magnetoconductivity ratio $[\sigma(B)-\sigma(0)]/\sigma(0)$ as a function of $B/T$. The magnetic field is parallel to the diamond surface. The data for different gate voltages and different temperatures collapse onto a single curve. (b) The data shown in (a) are plotted in the form of $\sigma(B)-\sigma(0)$ vs. $B/T$.}
\end{figure}

The fact that $[\sigma(B)-\sigma(0)]/\sigma(0)$ for different gate voltages follows a universal function of $B/T$ sets a strict constraint on a theory to account for this magnetoresistance effect. In the following, we consider several possible mechanisms to explain our findings. One might infer from the logarithmic increase of resistance with decreasing temperature shown in Fig. 1(b) that the two-dimensional weak localization or hole-hole interaction effects are important. A positive magnetoresistance is predicted by theories of these effects which take account of spin-orbit interaction and Zeeman splitting\cite{Hik80,Kaw81,Mae81,Lee82,Cas98}. In particular, interaction theories predict that $\sigma(B)-\sigma(0)$ is proportional to -$(B/T)^2$ at low magnetic fields\cite{Lee82,Cas98}. However, as its coefficient is completely independent of $\sigma(0)$\cite{Lee82,Bis82,Col02}, these theories cannot account for the observed $(B/T)^2$ dependence of $[\sigma(B)-\sigma(0)]/\sigma(0)$. Generally, theories of such perturbation corrections on the conductivity have a difficulty in accounting for the $B/T$ scaling of $[\sigma(B)-\sigma(0)]/\sigma(0)$. 

The in-plane large magnetoresistance effect strongly suggests that the Zeeman effect is important for the surface conductivity of hydrogen-terminated (100) diamond. Presumably, localized spins are present due to remaining dangling bonds and the carrier transport is affected by their magnetization, which is a function of $B/T$ if the interaction between the spins is negligible. An example of such a scenario is the prediction of a positive magnetoresistance in hopping transports\cite{Kur82,Mat95,Mei96,Agr98}: at sufficiently high magnetic fields, the hopping of an electron to the sites with a localized spin is suppressed due to the Pauli exclusion principle as the spin of the hopping electron and the localized spin are oriented in the same direction. In particular, the $(B/T)^2$ dependence of $[\sigma(B)-\sigma(0)]/\sigma(0)$ at low fields is predicted by a theory of nearest neighbor-hopping\cite{Agr98}. Although the temperature dependence of resistivity in our experiments is weaker than the exponential dependence expected for the hopping transport, the observed magnetoresistance may be understood along this line of reasoning. The fact that hydrogen-terminated (100) surface is reconstructed into ($2{\times}1$) surface having CH-CH dimer rows, while hydrogen-terminated (111) surface is not reconstructed\cite{Squ06}, indicates a larger density of remaining dangling bonds at the (100) surface. This may account for why this magnetoresistance effect is dominant for the (100) surface. In fact, the gate voltage dependence of the Hall carrier density at the (100) surface is weaker than that at the (111) surface\cite{Yam13}, which can be attributed to a larger trap density (dangling bond density) in the (100) surface. Another point that may be important is that due to the ionic liquid gating there is a large electric field ${\approx}10^{9}$ V/m perpendicular to the diamond surface, which should induce a Rashba spin-orbit coupling. This can lead to the anisotropy of the magnetoresistance effect.

Finally, it should be noted that Edmonds \textit{et al}. recently reported low-temperature magnetoresistance effects of hydrogen-terminated (100) diamond exposed to the air for several days\cite{Edm15}. An anomaly was observed at low magnetic fields ($\textless0.5$ T) at low temperatures (1.0-2.4 K), and it was interpreted as a weak antilocalization effect. However, we have not observed such an anomaly at $T{\ge}2.0$ K. In addition, a negative magnetoresistance was observed at 4.0 K in their study. The reason for these different behaviors is unclear at present, but it may be related to the different surface conditions, i.e., coating with ionic liquid or exposure to the air. 

In summary, the hole carriers accumulated at the hydrogen-terminated (100) surface of diamond shows a positive magnetoresistance, which contrasts with a negative magnetoresistance for the (111) surface. A large positive magnetoresistance appears even for the magnetic field parallel to the surface, indicating that the spin degree of freedom of the carriers plays an essential role in the surface conductivity. This magnetoresistance is presumably caused by the interactions between the spins of the carriers and localized spins arising from surface dangling bonds. We also find that the in-plane magnetoresistance ratio $[\rho(B)-\rho(0)]/\rho(0)$ for different gate voltages and for different samples follows a universal function of $B/T$. This scaling with $B/T$ cannot be fully accounted for, thus calling for the development of new theories. The observed spin-dependent transport may provide useful applications for diamond-based spintronics.

We acknowledge helpful discussions with S. Uji, Y. Ootuka, W. Izumida and T. Kato. We also thank T. Uchihashi for a critical reading of the manuscript. This study was supported by ALCA of JST, FIRST and KAKENHI (Grant No. 25287093) of JSPS, and "Nanotechnology Platform Project" of MEXT, Japan.

\end{document}